\documentclass[aps,prl,twocolumn,notitlepage,floatfix]{revtex4-2}

\usepackage{bbold}
\usepackage{lipsum}
\usepackage{amsmath,graphicx,color}
\usepackage{bm}
\usepackage{natbib}
\usepackage{placeins}
\setcitestyle{square}
\usepackage{braket}

\newcommand{\FSTx}{FeSe$_{1-x}$Te$_{x}$ }
\newcommand{\FST}{FeSe$_{0.45}$Te$_{0.55}$ }
\newcommand{\FSTS}{FeSe$_{0.3}$Te$_{0.7}$ }
\newcommand{\FSTT}{FeSe$_{0.45}$Te$_{0.55}$}

\begin{document}

\title{Competing topological superconducting phases in FeSe$_{0.45}$Te$_{0.55}$}

\author{Chang Xu, Ka Ho Wong, Eric Mascot, and Dirk K. Morr}
\affiliation{Department of Physics, University of Illinois at Chicago, Chicago, IL 60607, USA}

\begin{abstract}
    We demonstrate that recent angle-resolved photo-emission spectroscopy experiments provide strong evidence for the existence of two competing topological superconducting phases in FeSe$_{0.45}$Te$_{0.55}$. The coupling of their underlying microscopic mechanisms -- one based on a three-dimensional topological insulator, one based on two-dimensional superconductivity -- stabilizes topological superconductivity over a wide range of parameters, and gives rise to two disjoint topological regions in the phase diagram of FeSe$_{0.45}$Te$_{0.55}$. We show that the topological origin of these regions can be identified by considering the form of Majorana edge modes at domain walls.

\end{abstract}

\maketitle

{\it Introduction} Topological superconductors harbor Majorana zero modes whose non-Abelian braiding statistics and robustness against disorder and decoherence provide a new platform for the implementation of topological quantum computing \cite{Nayak2008}. Over the last few years, strong evidence for the existence of topological surface superconductivity in the iron-based superconductor \FST has emerged, ranging from the observation of a surface Dirac cone \cite{Zhang2018,Rameau2019,Zaki2019,Yangmu2021}, to that of Majorana zero modes (MZMs) in vortex cores, \cite{Wang2018,Machida2019,Kong2019,Zhu2020} and of Majorana edge modes at domain walls \cite{Wang2020}. However, the microscopic origin of these topological features has remained unclear. While they were originally attributed \cite{Wang2015,Wu2016,Xu2016,Zhang2018} to topological surface superconductivity arising from a Fu-Kane like mechanism \cite{Fu2008} of proximity induced superconductivity in the surface Dirac cone of a three-dimensional topological insulator -- referred to as the 3DTI mechanism --, recent experiments have cast doubt on this interpretation. In particular, angle-resolved photoemission spectroscopy (ARPES) experiments \cite{Rameau2019,Zaki2019,Yangmu2021} on FeSe$_{1-x}$Te$_{x}$ and quantum sensing experiments on \FSTS \cite{McLaughlin2021}, reported evidence for surface ferromagnetism \cite{Zaki2019} which can readily destroy the 3DTI mechanism \cite{Wu2021}. An alternative scenario \cite{Mascot2022,Wong2022} -- the 2DTSC mechanism -- was therefore proposed in which the two-dimensional (2D) nature of superconductivity in the $\alpha$-, $\beta$- and $\gamma$-bands of \FST \cite{Eschrig2009,Watson2015,Watson2016,Kreisel2020}, in combination with the observed surface magnetism \cite{Rameau2019,Zaki2019,Yangmu2021,McLaughlin2021} induces topological superconductivity. Interestingly enough, the experimentally observed opening of a gap at $E_F$ in the Dirac cone below $T_c$ \cite{Zaki2019} -- reflecting proximity induced superconductivity -- implies a coupling of these two competing mechanisms. The question thus naturally arises of how  the interplay between the competing 2DTSC and 3DTI mechanisms determines the topological properties and phase diagram of \FSTT.\par

In this Letter, we address this question and demonstrate that the competition between these two mechanism can not only explain the experimentally observed opening of two gaps in the surface Dirac cone of \FST -- at $E_F$ and at the Dirac point \cite{Zaki2019} -- but also gives rise to two disjoint topological regions in the phase diagram: a weak-moment region in which topological superconductivity arises from the 3DTI mechanism, and a large-moment region whose topological properties are determined by the 2DTSC mechanism. We demonstrate that the topological nature of these regions can be unambiguously identified by considering the electronic structure and currents near spin and $\pi$-phase domain walls. Our results provide unique characteristics allowing future experiments to elucidate the microscopic origin of the topological superconducting phases of \FSTT. \par

\begin{figure}[bt]
    \centering
    \includegraphics[width=8.5cm]{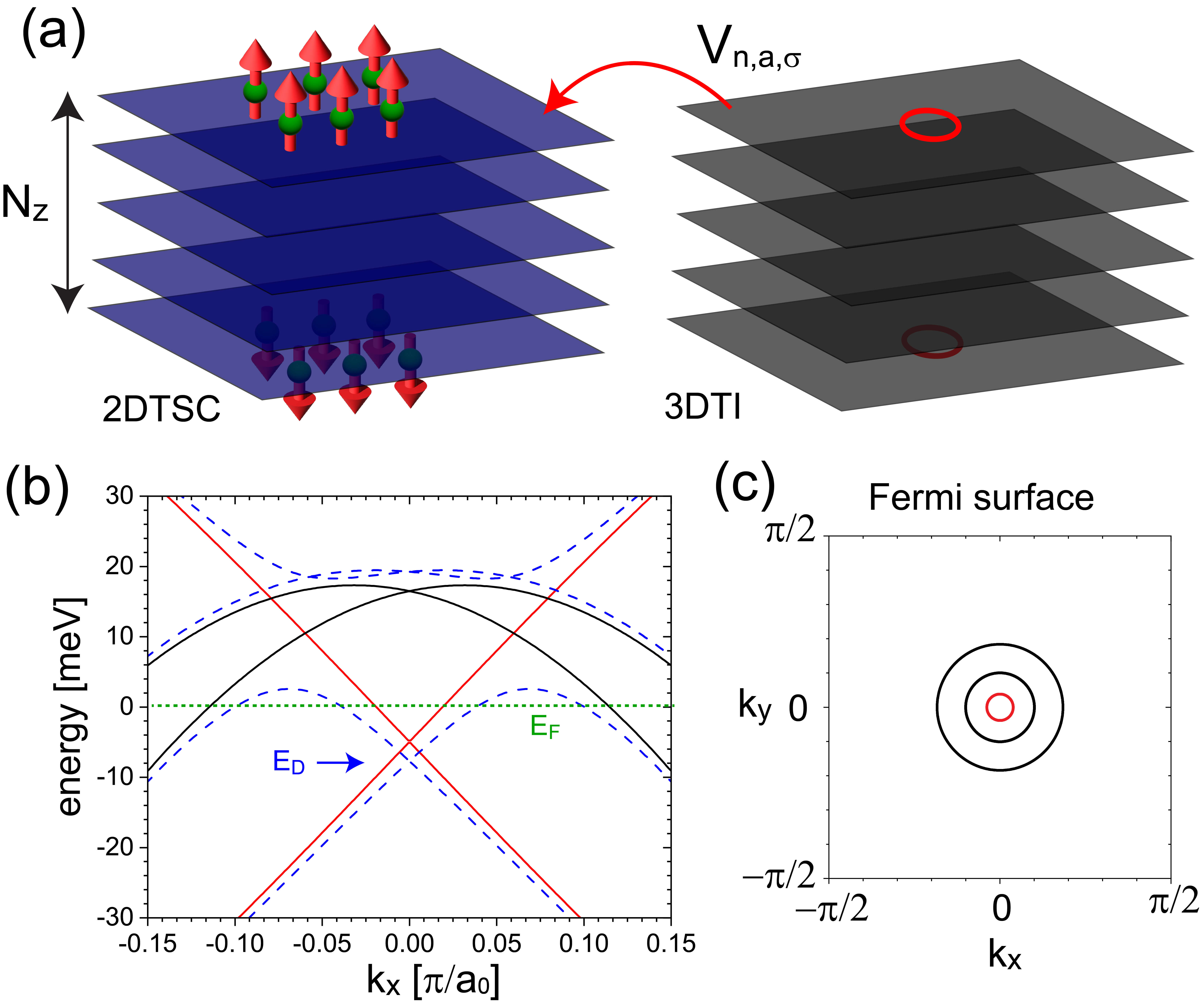}
    \caption{(a) Schematic representation of the coupled 3DTI and 2DTSC systems with $N_z$ layers. (b) Surface electronic structure: solid red and black lines represent the decoupled 3DTI and 2DTSC ($V = 0$) systems, respectively, dashed lines the coupled ones ($V \not = 0$). (c) Fermi surface of the coupled system. Parameters are $(JS,\alpha,\mu, \mu_c)=(0,0.1,3.8,0.06)t$.}
    \label{fig:fig1}
\end{figure}

{\it Theoretical Formalism}
To study the emergence of topological surface superconductivity in \FSTT, we consider a three-dimensional (3D) system with $N_z$ layers [see Fig.~\ref{fig:fig1}(a)]. To meet the resulting computational demands, we utilize a simplified version of the 2DTSC model \cite{Mascot2022}, which nevertheless preserves its salient features [see Supplemental Material (SM) Sec.I]. In particular, it reproduces the experimentally observed \cite{Borisenko2016} two Fermi surfaces that are closed around the $\Gamma$-point, which are relevant for the coupling to the Dirac cone. Moreover, the quasi two-dimensional (2D) nature of superconductivity in \FSTx, as observed by ARPES experiments \cite{Watson2015,Watson2016,Kreisel2020} implies zero direct coupling between the 2DTSC, yielding the Hamiltonian
\begin{align}
H_{2DTSC}&= - t\sum_{\langle \mathbf{rr'}\rangle,\sigma,n} f_{n,\mathbf{r},\sigma}^{\dagger} f_{n,\mathbf{r'},\sigma}\ -\mu \sum_{\mathbf{r},\sigma} f_{n,\mathbf{r},\sigma}^{\dagger} f_{n,\mathbf{r},\sigma} \nonumber \\
&+  \Delta_0 \sum_{n,\mathbf{r}} f_{n,\mathbf{r}, \uparrow}^{\dagger} f_{n,\mathbf{r}, \downarrow}^{\dagger} + H.c. \nonumber \\
&+  \sum_{ n=1,N_z } g_n \left( i \alpha \sum_{{\bf r}, \boldsymbol{\delta}, \sigma, \sigma^\prime} f^\dagger_{n,{\bf r}, \sigma}  \left[\boldsymbol{\delta} \times \boldsymbol{\sigma} \right]^z_{\sigma, \sigma^\prime}   f_{n,{\bf r + \boldsymbol{\delta}}, \sigma^\prime} \right. \nonumber \\
& + \left. JS \sum_{\mathbf{r},\sigma,\sigma'} f_{n,\mathbf{r},\sigma}^{\dagger}\sigma_{\sigma \sigma'}^{z} f_{n,\mathbf{r},\sigma'} \right) \ ,
\end{align}
where $f^\dagger_{n,{\bf r},\sigma}$ creates an electron with spin $\sigma$ at site ${\bf r}$ in layer $n=1,..,N_z$, $-t$ is the electronic hopping amplitude between nearest-neighbor sites on a 2D square lattice, $\mu$ is the chemical potential, $\Delta_0$ is the $s$-wave superconducting order parameter, $\alpha$ is the Rashba spin-orbit coupling, $J$ is the magnetic exchange coupling, and $S$ is the ordered moment. We take the experimentally observed out-of-plane magnetism and  Rashba spin-orbit interaction arising from a broken inversion symmetry to be present on the top and bottom surfaces only, with $g_{1}=-g_{N_z}=1$, as required by symmetry (see SM Sec.I).

In contrast, the 3DTI mechanism assumes that the bulk bands of \FSTx represent a three-dimensional topological insulator, described by the Hamiltonian \cite{Schubert2012}
\begin{align}
H_{3DTI}= & \sum_{{\bf r},j=1,2,3} \left( \Psi_{{\bf r}+\hat{e}_j}^{\dagger}  \frac{-t\Gamma^1-i\lambda \Gamma^{j+1}}{2} \Psi_{\bf r} + H.c. \right) \nonumber \\
+& \sum_{\bf r} \Psi_{\bf r}^\dagger (\mu_c \Gamma^0 + m \Gamma^1) \Psi_{\bf r}
\end{align}
with spinor $ \Psi_{{\bf r}}^{\dagger} = \left( c_{{\bf r},1,\uparrow}^{\dagger}, c_{{\bf r},2,\uparrow}^{\dagger}, c_{{\bf r},1,\downarrow}^{\dagger}, c_{{\bf r},2,\downarrow}^{\dagger} \right)$
where $c_{{\bf r},a,\sigma}$ annihilates an electron with spin $\sigma$ in orbital $a=1,2$ at site ${\bf r}$,  $\Gamma^{0,...,4} = (\mathbb{1} \otimes \mathbb{1}, \mathbb{1} \otimes s_z, -\sigma_y \otimes s_x, \sigma_x \otimes s_x, -\mathbb{1} \otimes s_y)$ with $\sigma_i$ and $s_i (i=x,y,z)$ being Pauli matrices, $\mu_c$ is the chemical potential, and $\lambda$ is the spin-orbit coupling. We take $m=2t$ and $\lambda=t$ such that the system is in the topological phase \cite{Schubert2012}, and consider a 3DTI with $N_z$ layers. The coupling between the 3DTI and 2DTSC layers is described by the Hamiltonian
\begin{align}
    H_{hyb} =& - \sum_{n,\mathbf{r},a,\sigma}  V_{n,a,\sigma} f_{n,\mathbf{r},\sigma}^{\dagger} c_{n,\mathbf{r},a,\sigma}  + H.c. \ ,
\end{align}
with $V_{n,a,\sigma}$ being the hybridization strength in the $n$'th layer. The relative signs of $V_{n,a,\sigma}$ are determined by the rotation symmetry of the coupled system (see SM Sec.II). \par

{\it Results}
In Fig.~\ref{fig:fig1}(b), we present the electronic dispersion of the decoupled ($V=0$) and coupled ($V\not =0$) 3DTI and the 2DTSC systems above $T_c$ where $JS=0$. The dispersion exhibits a Dirac cone with a Dirac point located at $E_D$ (here, we scaled $t$ such that $E_D \approx -8$meV, as experimentally observed \cite{Zaki2019}) and two hole-like bands of the 2DTSC, which are split by the Rashba spin-orbit interaction \cite{Borisenko2016}. For $V \not = 0$, the 3DTI and 2DTSC bands hybridize, thus opening a hybridization gap at the band crossings. The resulting Fermi surfaces shown in Fig.~\ref{fig:fig1}(c), with the (red) innermost one arising predominantly from the Dirac cone, and the two outer (black) ones due to the 2DTSC bands, are in qualitative agreements with the experimental ARPES observations \cite{Zhang2018}.
\begin{figure}[bt]
    \centering
    \includegraphics[width=8cm]{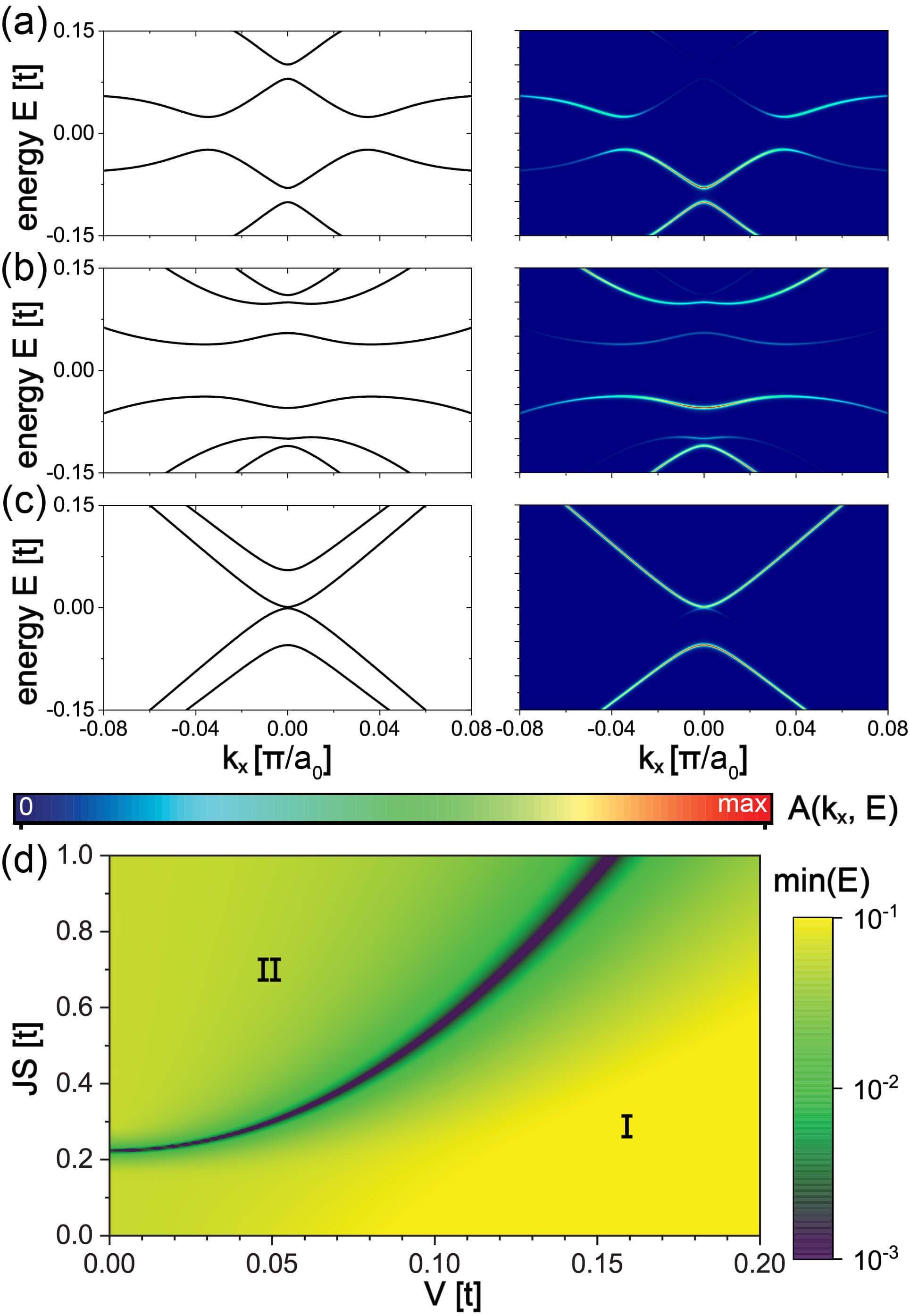}
    \caption{Evolution of the electronic structure (left column) and spectral functions (right column, for $k_y=0$) below $T_c$ with increasing $JS$: (a) $(JS,\mu_c)=(0.1,0.06)t$ (b) $(JS,\mu_c)=(0.2,0.04)t$, and (c) $(JS,\mu_c)=(0.45,0.04)t$. (d) Gap at the $\Gamma$-point as a function of $V$ and $JS$. Parameters are $(V,\Delta_0,\alpha,\mu)=(0.07,0.1,0.1,3.8)t$ and $N_z=5.$
    }
    \label{fig:fig2}
\end{figure}
To describe the experimentally observed temperature evolution of the spectral function below $T_c$ \cite{Zaki2019}, and specifically, the opening of gaps at the Fermi energy and at the Dirac point, we assume an onset of the magnetic order ($JS \not = 0$) at $T_c$, with $JS$ increasing with decreasing temperature. The resulting evolution of the electronic structure and (3DTI) $c$-electron surface spectral functions is shown in Fig.~\ref{fig:fig2} (the $f$-electron spectral function is shown in SM Sec.III). Just below $T_c$, [see Fig.~\ref{fig:fig2}(a)], the hybridization between the 2DTSC and the 3DTI system proximity induces a superconducting gap, $\Delta_{SC}$, in the 3DTI Dirac cone at $E_F$. At the same time, a non-zero magnetization $JS \not = 0$ leads to a magnetic polarization in the Dirac cone, opening a gap, $\Delta_D$, at the Dirac point. As we demonstrate below, this renders the system a Fu-Kane type topological superconductor \cite{Fu2008}. With increasing $JS$, both $\Delta_{SC}$ and $\Delta_D$ further increase [see Fig.~\ref{fig:fig2}(b)], in good qualitative agreement with the temperature dependence of the spectral function observed in ARPES experiments \cite{Rameau2019,Zaki2019,Yangmu2021} (see, e.g., Fig.4 in Ref.\cite{Zaki2019}). Eventually, $\Delta_{SC}$ closes at the $\Gamma$-point [see Fig.~\ref{fig:fig2}(c)] at $(JS)_{pt}$, which is a general feature of the coupled system and part of a line of gap closings (occurring only at the $\Gamma$ point, see SM Sec.IV), resulting in two disjoint regions in the $(V,JS)$-plane [see Fig.~\ref{fig:fig2}(d)]. This naturally raises the question of whether this gap closing represents a topological phase transition, and if so, what the topological nature of the involved phases in regions I and II are. To answer this question, we first note that the gap closing occurs only in the surface spectral function, while the bulk remains gapped, implying that it is associated with a transition affecting the topological nature of the surface phase. Moreover, we can consider two limiting cases: at $JS=0$ and $V \not = 0$, the proximity induced superconducting gap in the 3DTI Dirac cone is expected to lead to Fu-Kane type \cite{Fu2008} topological surface superconductivity, which should thus hold for the entire region I. In contrast, for $V \rightarrow 0$, the gap closing line terminates at a value of $(JS)_{pt}$, such that for $JS>(JS)_{pt}$, the 2DTSC is topological, while for $JS<(JS)_{pt}$ it is trivial. This suggests that in region II, the system exhibits topological surface superconductivity arising from the 2DTSC mechanism, and that the gap closing line thus indeed represents a topological phase transition. However, as the spectral weight in the negative energy branch of the band in which the gap closing occurs is vanishingly small [see Fig.~\ref{fig:fig2}(c)], the gap closings might be difficult to detect in ARPES experiments \cite{Sobota2021}. To further elucidate the topological nature of regions I and II, we next consider their electronic structure near vortices and domain walls.\par

{\it MZMs in a Vortex core}
\begin{figure}[bt]
    \centering
    \includegraphics[width=8.5cm]{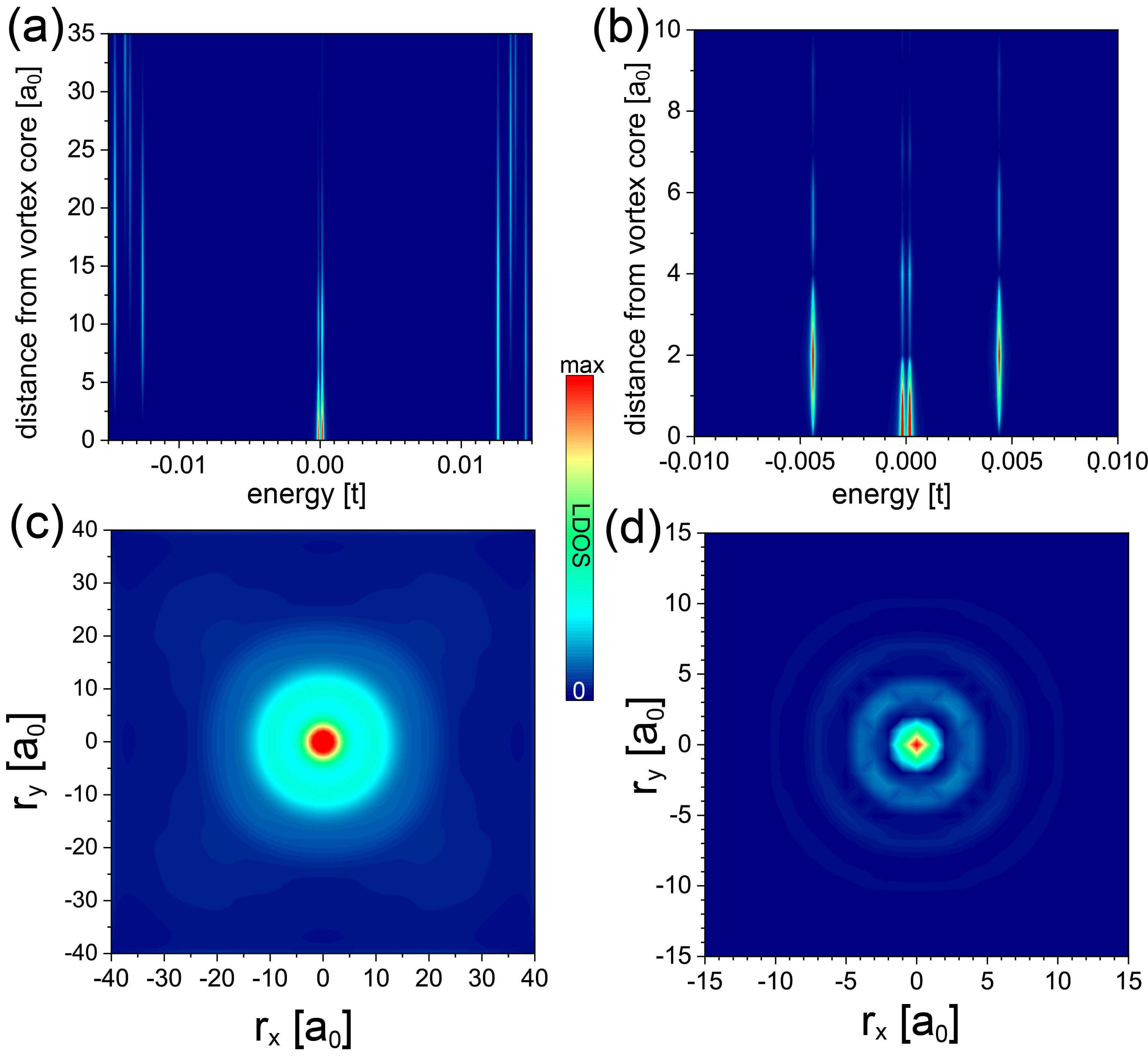}
    \caption{LDOS as a function of energy and distance from the vortex core on the top surface for (a) the $c$-electrons in region I, and (b) the $f$-electrons in region II (for details, see SM Sec. VI). (c), (d) Spatial plot of the zero-energy LDOS corresponding to (a) and (b), respectively.}
    \label{fig:fig3}
\end{figure}
To explore the emergence of Majorana zero modes in magnetic vortices,  we consider a system with a finite extent in the $x-,y-$, and $z-$direction, and model a vortex \cite{Wu2021} by assigning a phase to the local superconducting order parameter, $\Delta({\bf r}) = |\Delta_0| e^{i \phi({\bf r})}$, where $\phi({\bf r})$ is a position dependent angle (for details see SM Secs.V and VI).
\begin{figure*}[htb]
    \centering
    \includegraphics[width=\textwidth]{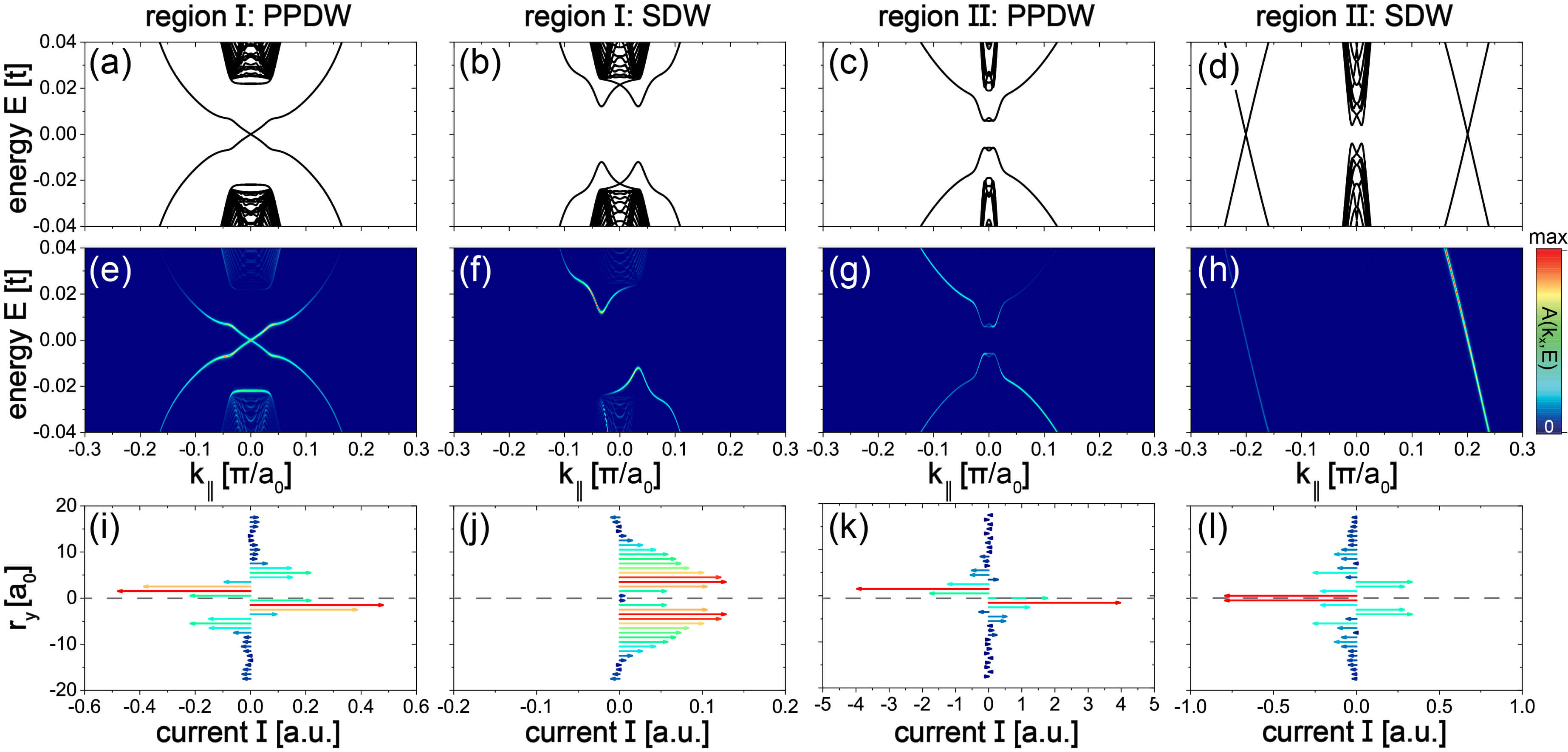}
    \caption{(a) - (d) Electronic structures as a function of momentum, $k_\parallel$ along a PPDW and SDW domain wall, for parameter sets characteristic of regions I and II (see SM Sec.VII). Corresponding spectral functions of the (e),(f)  $c$-electrons, and  (g),(h) $f$-electrons. (i) - (l) Currents along the domain walls (denoted by a dashed gray line).}
    \label{fig:fig4}
\end{figure*}
In Figs.~\ref{fig:fig3}(a),(b), we present the zero-energy LDOS along a line-cut through the vortex core in regions I and II of the phase diagram, respectively. In both cases, the LDOS exhibits a low-energy state at $E=\pm \varepsilon$, whose spatial structure at $+ \varepsilon$ is shown in Figs.~\ref{fig:fig3}(c),(d), respectively. The localization of the LDOS at the site of the vortex core, together with $\varepsilon \rightarrow 0$ with increasing system size, implies that this (near) zero energy state is an MZM, and that in both regions, the superconducting surface phase is topological in nature. However, the existence of MZMs in both regions also implies that their experimental observation cannot discriminate between these two regions; this, however, can be achieved by considering the electronic structure near domains walls.\par

{\it Majorana edge modes along domain walls}
The emergence of Majorana edge modes at certain types of domain walls was shown to provide insight into the microscopic origin underlying topological superconductivity \cite{Mascot2022}. In particular, in a 2DTSC, a chiral Majorana edge mode emerges at a spin domain wall (SDW), where the magnetization flips its orientation, ${\bf S} \rightarrow -{\bf S}$, but not a $\pi$-phase domain wall (PPDW) where the superconducting order parameter undergoes a sign change, $\Delta_0 \rightarrow -\Delta_0$ \cite{Mascot2022}. In contrast, in a Fu-Kane type topological superconductor, a Majorana edge mode emerges at a PPDW only \cite{Fu2008}. To test whether this qualitative difference allows us to identify the topological nature of regions I and II, we present in Figs.~\ref{fig:fig4}(a)-(d) the electronic dispersion along a domain wall for a PPDW and SDW for representative parameter sets in both regions. In region I, only the PPDW exhibits an in-gap mode that traverses the superconducting gap [Fig.~\ref{fig:fig4}(a)], which together with its robustness against disorder effects (see SM Sec.VII), identifies it as a Majorana edge mode. In contrast, the SDW only possesses trivial in-gap states [Fig.~\ref{fig:fig4}(b)], implying that topological surface superconductivity in region I arises from the Fu-Kane-like 3DTI mechanism \cite{Fu2008}. Conversely, in region II, only a SDW exhibits a Majorana edge mode [Fig.~\ref{fig:fig4}(d)], which is unaffected by disorder (see SM Sec.VII), while the PPDW does not [Fig.~\ref{fig:fig4}(c)].  Thus, in region II,  topological superconductivity arises from the 2DTSC mechanism. Moreover, the complementary emergence of Majorana edge modes along a PPDW and SDW in regions I and II also implies that the topological phases arising from these two mechanisms are mutually exclusive and thus competing, rather than coexisting. Moreover, a plot of the spectral functions at the domain walls  [see Figs.~\ref{fig:fig4}(e)-(h)] reveals that the Majorana mode along the SDW in region II is chiral in nature [cf. Fig.~\ref{fig:fig4}(d) and (h)] (see SM Sec.VII), while that of the PPDW in region I is neither helical nor chiral. This qualitative difference can be detected using quasiparticle interference spectroscopy \cite{Hoffman2002}, as the parallel Majorana branches in Fig.~\ref{fig:fig4}(h) lead to a nearly  dispersion-less peak in the QPI spectrum \cite{Chang2023}.

The observation of a Majorana edge mode along a domain wall is in general not sufficient to identify the underlying microscopic origin, unless the nature of the domain wall is known. The latter can be achieved by considering the screening currents in the vicinity of domain walls (see SM Sec.VIII), as shown in Figs.~\ref{fig:fig4}(i)-(l). While both the PPDW and SDW induce screening currents, the resulting current pattern is qualitatively different. Since the Chern number, and hence the chirality, is reversed at a SDW, the currents on both sides of the domain wall are symmetric, leading to a non-zero net current along the domain wall. In contrast, the current patterns on both sides of a PPDW are antisymmetric, and the net current is thus zero. Since the net current along a domain wall can be measured using a superconducting quantum interference device \cite{Spanton2014}, its presence or absence is a crucial feature distinguishing a SDW from a PPDW. Thus, the presence or absence of a Majorana edge mode together with that of a net current allows one to unambiguously identify the nature of the domain wall, and thus the origin of the underlying topological phase. \par

{\it Conclusions} We demonstrated that the opening of two gaps in the surface Dirac cone of \FSTT, as reported by recent ARPES experiments \cite{Zaki2019}, provides strong evidence for the existence of two competing mechanisms underlying the emergence of  topological superconductivity. The competition between these mechanisms -- the 2DTSC and 3DTI mechanisms -- while giving rise to robust topological surface superconductivity over a large range of parameters, also produces two disjoint topological regions in parameter space. By considering the emergence of Majorana edge modes at a SDW and PPDW, we showed that topological superconductivity in region I arises from the 3DTI mechanism, while that in region II is due to the 2DTSC mechanism.
An important outstanding question remains which mechanism is responsible for the topological features of \FSTT, such as vortex core MZMs \cite{Machida2019}, observed at mK temperatures. While the experimental ARPES observations \cite{Zaki2019} together with our results in Fig.~\ref{fig:fig2} suggest that topological superconductivity just below $T_c$ arises from the 3DTI mechanism, they also show that with decreasing temperature, the system approaches, and potentially even crosses the topological phase transition into region II; a transition which might be difficult to observe via ARPES due to the vanishingly small spectral weight in the gap closing bands. Clearly, future experiments are required to elucidate the nature of the topological phase in \FST at the lowest temperatures. \\

\begin{acknowledgments}
The authors would like to thank P. Johnson and S. Rachel for stimulating discussions. This work was supported by the U. S. Department of Energy, Office of Science, Basic Energy Sciences, under Award No. DE-FG02-05ER46225.
\end{acknowledgments}

\end{document}


\preprint{APS/123-QED}


\title{Competing topological superconducting phases in \FSTT \\[0.25cm]
{\large Supplementary Material}}

\author{Chang Xu, Ka Ho Wong, Eric Mascot, and Dirk K. Morr}
\affiliation{Department of Physics, University of Illinois at Chicago, Chicago, IL 60607, USA}

\maketitle

\section{The 2DTSC and 3DTI Hamiltonians}

To study the emergence of topological surface superconductivity in \FSTT, we consider a three-dimensional (3D) system with $N_z$ layers [see Fig.~1(a) in the main text]. The 2DTSC model previously introduced \cite{Mascot2022} to describe the emergence of topological surface superconductivity in \FSTT, starts from the quasi two-dimensional (2D) nature of superconductivity in \FSTxx, utilizing a 5-band model that was shown to explain the emergence of a superconducting phase with $s_\pm$-symmetry in the $\alpha$-, $\beta$- and $\gamma$-bands, which arise predominantly from the Fe d-orbital \cite{Eschrig2009,Watson2015,Watson2016,Kreisel2020}. These bands
exhibit Fermi surfaces that are closed around the $\Gamma$ and $X/Y$-points in the 1Fe Brillouin zone. However, as we need to consider a 3D systems with $N_z$ layers (in order to obtain Dirac cones on the 3DTI surfaces), the inclusion of the full 5-band model is computationally prohibitive. We therefore consider a simplified version of the 2DTSC model \cite{Mascot2022}, as given in Eq.(1) of the main text, which nevertheless preserves the salient features of the full 5-band model. In particular, it reproduces the experimentally observed \cite{Borisenko2016} two Fermi surfaces that are closed around the $\Gamma$-point, which are relevant for the coupling to the Dirac cone, as shown in Fig.1(b) of the main text.

Moreover, the magnetic nature of Fe suggest that the experimentally observed magnetism on the surface of \FST \cite{Rameau2019,Zaki2019,Yangmu2021} arises from the Fe d-orbitals, rather than the Te/Se orbitals, or the hybridized Te $p_z$ and Fe $d_{xz}$-orbitals that give rise to the 3DTI \cite{Wang2015,Wu2016,Xu2016,Zhang2018}. We therefore included the effect of the experimentally observed magnetization into the 2DTSC Hamiltonian, as previously discussed \cite{Mascot2022}. As a result, not only superconductivity, but also a magnetization is proximity induced into the 3DTI.  Finally, we note that rotational symmetry of the system requires that the magnetization and Rashba spin-orbit interaction possess opposite signs on the top and bottom surfaces, i.e., $J \rightarrow -J$ and $\alpha \rightarrow -\alpha$.

\begin{figure*}[htbp]
    \centering
    \includegraphics[width=6 in]{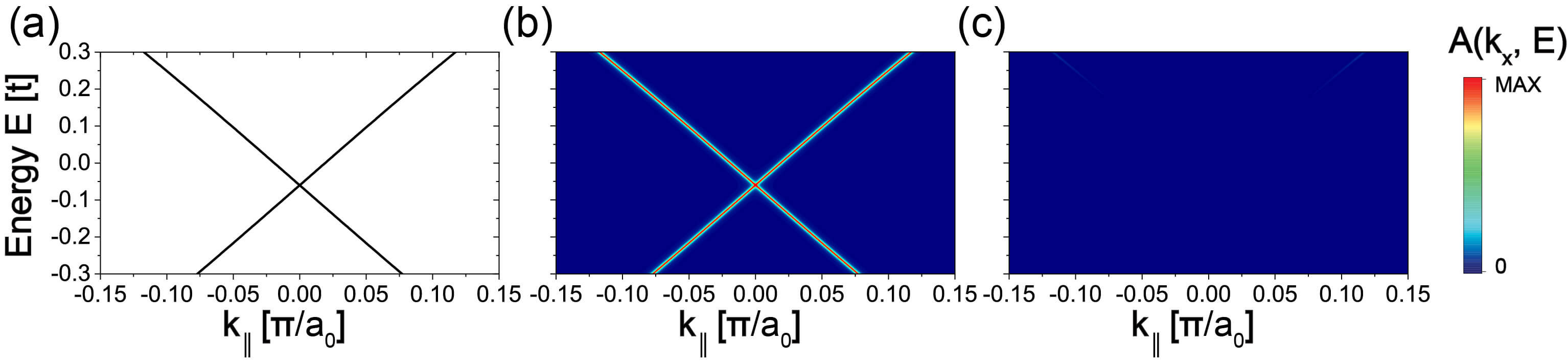}
    \caption{(a) Surface electronic structure of the 3DTI ($\mu_c=0.06$) in normal state, revealing a Dirac cone. Spectral function of the (3DTI) $c$-electrons (b) on the surface layer, and (c) on the first layer below the surface.}
    \label{SIfig:fig1}
\end{figure*}
Moreover, for the calculations presented in the main text, we consider a 3DTI with a finite number of layers in the $z$-direction, $N_z$. As a result, a Dirac cone appears on the top and bottom surfaces of the system. To demonstrate this, we present in Fig.~\ref{SIfig:fig1}(a) the electronic structure of the 3DTI only ($V=0$) in the normal state, which exhibits the characteristic Dirac cone. A comparison of the (3DTI) $c$-electron spectral function on the surface layer [see Fig.~\ref{SIfig:fig1}(b)], and in the first layer below the surface [see Fig.~\ref{SIfig:fig1}(c)], demonstrates that the Dirac cone exists only on the surface of the 3DTI, and essentially possesses no spectral weight in the layers below the surface.

\section{Derivation of the hybridization matrices between the 2DTSC and 3DTI systems}

To derive the form of the hybridization elements, $V_{n,a,\sigma}$ [see Eq.(3) of the main text], we use that the coupled 2DTSC and 3DTI system, with a finite number of $N_z$ layers in the $z$-direction, is invariant under rotation around the $x$-axis. In the following, we first consider the symmetry properties of the 3DTI and 2DTSC separately, followed by a symmetry study of the coupled system.

\subsection{Rotation symmetry of the 3DTI}

The Hamiltonian of the 3DTI \cite{Schubert2012} in Eq.(2) of the main text can be written as $H_{3DTI} = \sum_{\bf k} \Psi_{{\bf k}}^{\dagger} {\cal {\hat H}}({\bf k},\lambda) \Psi_{{\bf k}} $ with spinor $\Psi_{{\bf k}}^{\dagger} = \left( c_{{\bf k},1,\uparrow}^{\dagger}, c_{{\bf k},2,\uparrow}^{\dagger}, c_{{\bf k},1,\downarrow}^{\dagger}, c_{{\bf k},2,\downarrow}^{\dagger} \right)$ and
\begin{align}
   {\cal  {\hat H}}({\bf k},\lambda) &=\left(\begin{array}{cccc}
         -\mu_c + \xi_{\bf k}& i\lambda\sin(k_z)& 0& \lambda[i\sin(k_x)+\sin(k_y)]\\
         -i\lambda\sin(k_z)&-\mu_c -\xi_{\bf k}& \lambda[i\sin(k_x)+\sin(k_y)]&0\\
         0&\lambda[-i\sin(k_x)+\sin(k_y)]&-\mu_c +\xi_{\bf k}&i\lambda \sin(k_z)\\
         \lambda[-i\sin(k_x)+\sin(k_y)]& 0& -i\lambda \sin(k_z)&-\mu_c -\xi_{\bf k}
    \end{array}\right)
\end{align}
where $\xi_{\bf k}=m-2t[\cos(k_x) + cos(k_y) + cos(k_z)]$. The Hamiltonian is invariant under a simultaneous rotation around the $x$-axis, both in real and spin space, and $\lambda \rightarrow -\lambda$, which is achieved using the unitary transformation
\begin{align}
    {\hat M}_x = \sigma_x \otimes \sigma_z
\end{align}
with $\sigma_i$ being the Pauli matrices, such that
\begin{align}
    {\hat M}_x {\cal {\hat H}}(k_x,-k_y,-k_z,\lambda) {\hat M}^\dagger_x = {\cal  {\hat H}}(k_x,k_y,k_z,\lambda)
    \label{eq:TIsymm}
\end{align}

For the system considered in the main text, the 3DTI consists of a finite number of $N_z$ layers in the $z$-direction. Moreover, as superconductivity is proximitized into the 3DTI, we need to write the Hamiltonian for the 3DTI with $N_z$ layers in Nambu space, resulting in
\begin{align}
   {\cal {\hat  H}}[k_x, k_y, i=(1,...,N_z),\lambda]  =
\left( \begin{array}{ccccc}
{\cal {\hat H}}_{1}      &  {\hat T}_z         &      0 &  0 & 0         \\
{\hat T}_z^{\dagger}   & {\cal {\hat H}}_{2}          &   {\hat T}_z      &  0 & 0  \\
0 & \ddots & \ddots & \ddots & 0 \\
0 & 0 & {\hat T}_z^{\dagger} & {\cal {\hat H}}_{N_z-1} & {\hat T}_z \\
         0 & 0 & 0       &{\hat T}_z^{\dagger}  &  {\cal {\hat H}}_{N_z}
\end{array} \right)
\end{align}
with spinor $ \Psi_{{\bf k}}^{\dagger} = \left( \Psi_{1,{\bf k}}^{\dagger}, \Psi_{2,{\bf k}}^{\dagger}, ..., \Psi_{N_z,{\bf k}}^{\dagger}\right)$ and $\Psi_{i,{\bf k}}^{\dagger} = \left( c_{i,{\bf k},1,\uparrow}^{\dagger}, c_{i,{\bf k},2,\uparrow}^{\dagger}, c_{i,{\bf k},1,\downarrow}^{\dagger}, c_{i,{\bf k},2,\downarrow}^{\dagger}, c_{i,{\bf -k},1,\uparrow}, c_{i,{\bf -k},2,\uparrow}, c_{i,{\bf -k},2,\downarrow}, c_{i,{\bf -k},2,\downarrow} \right)$, and $i$ being the layer index. Moreover, ${\cal {\hat H}}_{i}$ is Hamiltonian in Nambu space of the $i$'th layer of the 3DTI  given by
\begin{align}
   {\cal {\hat H}}_{i}(k_x, k_y) & = \left( \begin{array}{cc}
{\cal {\hat H}}^n_i(k_x, k_y)      &   0         \\
 0 & -\left[ {\cal {\hat H}}^n_i(-k_x, -k_y) \right]^T
\end{array} \right)
\end{align}
where ${\cal {\hat H}}^n_i$ is the normal state Hamiltonian of the $i$'th layer of the 3DTI given by
\begin{align}
    {\cal {\hat H}}^n_i & = \left(\begin{array}{cccc}
         -\mu_c + \varepsilon_{\bf k}& 0 & 0& \lambda[i\sin(k_x)+\sin(k_y)]\\
         0 &-\mu_c -\varepsilon_{\bf k}& \lambda[i\sin(k_x)+\sin(k_y)]&0\\
         0&\lambda[-i\sin(k_x)+\sin(k_y)]&-\mu_c +\varepsilon_{\bf k}& 0\\
         \lambda[-i\sin(k_x)+\sin(k_y)]& 0& 0 &-\mu_c -\varepsilon_{\bf k}
    \end{array}\right)
\end{align}
with $\varepsilon_{\bf k}=m-2t[\cos(k_x) + cos(k_y)]$. The hopping matrix ${\hat T}_z$ is given by
\begin{align}
    {\hat T}_z = \left(\begin{array}{cc}
{\hat t}_z      &   0         \\
 0 & -{\hat t}_z^T
\end{array} \right)
\end{align}
where
\begin{align}
    {\hat t}_z &= \left(\begin{array}{cccc}
         t&-\lambda&0&0\\
         \lambda&-t&0&0\\
         0&0&t&-\lambda\\
         0&0&\lambda&-t
    \end{array}\right) \ .
\end{align}

Defining next
\begin{align}
    {\hat S}_{x} =
\left( \begin{array}{ccccc}
   0    & 0 &  \cdots    & 0 & {\hat N}_{x} \\
   0     & \cdots   &   0 & {\hat N}_{x} & 0  \\
   \vdots    &  \vdots & \ddots & \vdots & \vdots \\
   0 & {\hat N}_{x} & 0 & \cdots & 0 \\
{\hat N}_{x}  &   0    &    \cdots & 0 & 0
\end{array} \right)
\end{align}
where
\begin{align}
    {\hat N}_{x} = \left(\begin{array}{cc}
{\hat M}_x      &   0         \\
 0 & {\hat M}_z
\end{array} \right) \ ,
\end{align}
the symmetry of Eq.(\ref{eq:TIsymm}) becomes
\begin{align}
    {\hat S}_{x}^{\dagger} {\cal {\hat H}}[k_x, -k_y, i=(N_z,...,1),\lambda] {\hat S}_{x} = {\cal {\hat H}}[k_x, k_y, i=(1,...,N_z),\lambda] \ .
\end{align}

\subsection{Rotation symmetry of the 2DTSC}

The Hamiltonian for a single layer of the 2DTSC can be written as $H_{TSC} = \sum_{\bf k} \Phi_{{\bf k}}^{\dagger} {\hat H}_{TSC}({\bf k},\alpha,J) \Phi_{{\bf k}} $ with spinor $\Phi_{{\bf k}}^{\dagger} = \left( f_{{\bf k},\uparrow}^{\dagger}, f_{{\bf k},\downarrow}^{\dagger}, f_{{\bf -k},\uparrow}, f_{{\bf -k},\downarrow} \right)$ and
\begin{align}
    {\hat H}_{TSC}({\bf k},\alpha,J) &=\left(\begin{array}{cccc}
          \xi_{\bf k}+J& 2\alpha[i \sin(k_x)+\sin(k_y)]& 0& \Delta\\
         2\alpha[-i\sin(k_x)+\sin(k_y)]& \xi_{\bf k}-J& -\Delta&0\\
         0&-\Delta&-\xi_{\bf k}-J&2\alpha[-i\sin(k_x)+\sin(k_y)]\\
         \Delta& 0&2\alpha[i\sin(k_x)+\sin(k_y)]& -\xi_{\bf k}+J
    \end{array}\right)
\end{align}
where $\epsilon_{\bf k}=-2 t[\cos(k_x) + \cos(k_y)]-\mu$. Next we define a unitary matrix ${\hat U}_x$ via
\begin{align}
    {\hat U}_x &= \sigma_z \otimes \sigma_x =
    \left(\begin{array}{cccc}
         0&1&0&0\\
         1&0&0&0\\
         0&0&0&-1\\
         0&0&-1& 0
    \end{array}\right)
\end{align}
which yields
\begin{align}
    {\hat U}_{x}^{\dagger} {\hat H}_{TSC}(k_x, -k_y, \Delta, -\alpha, -J) {\hat U}_{x} = {\hat H}_{TSC}(k_x, k_y, \Delta, \alpha, J)
\end{align}
As we show in the next section, this symmetry operation is required, as a rotation of the entire coupled system around the $x$-axis exchanges the top and bottom 2DTSC surfaces which possess an opposite sign in the magnetization and Rashba spin-orbit interaction.

\subsection{Rotation symmetry of the coupled 3DTI and 2DTSC systems}

We next consider the rotation symmetry of the coupled 3DTI and 2DTSC systems. To this end, we consider a 3DTI with $N_z$ layers, and only two layers of the 2DTSC system that are coupled to the top and bottom surfaces of the 3DTI system. As we discuss in the main text, and explicitly show in SM Sec. V, the bulk bands of the 2DTSC do not affect the electronic structure or topological phase on the surface of the system. Moreover, as the 2DTSC bulk layers do not contain a Rashba spin-orbit interaction, or ferromagnetic magnetization, their rotation properties are trivial, and we will therefore omit them below for clarity of the derivation. \\

The Hamiltonian of the 3DTI with $N_z$ layers and of the two 2DTSC layers is given by
\begin{align}
    &{\hat H}_S(k_x, k_y, N_z.\alpha, J, \lambda) =
\left( \begin{array}{cccccc}
{\hat H}_{TSC}(\alpha, J)  &{\hat V}_{t}     &       0      &     0         &       0        &     0      \\
{{\hat V}_{t}}^{\dagger} &{\cal {\hat H}}_1    &  {\hat T}_z   &        0       &       0        &    0       \\
       0     &{\hat T}_z^{\dagger}  &{\cal {\hat H}}_2    &{\hat T}_z          &      0         &    0       \\
        0    &      0         &\ddots  & \ddots        &\ddots            &      0     \\
        0    &       0        &     0          &{\hat T}_z^{\dagger}  &{\cal {\hat H}}_{N_z}  &{\hat V}_{b}         \\
        0    &       0        &       0        &     0          &{\hat V}_{b}^{\dagger}  &{\hat H}_{TSC}(-\alpha, -J)
\end{array} \right)
\label{eq:coupledH}
\end{align}
with spinor $ \Psi_{{\bf k}}^{\dagger} = \left( \Phi_{1,{\bf k}}^{\dagger}, \Psi_{1,{\bf k}}^{\dagger}, \Psi_{2,{\bf k},}^{\dagger}, ..., \Psi_{N_z,{\bf k}}^{\dagger}, \Phi_{N_z,{\bf k}}^{\dagger}\right)$, and ${\hat V}_{t,b}$ are the hybridization matrices between the 2DTSC layers and the top and bottom 3DTI surface layers, as described by Eq.(3) of the main text.
Using next
\begin{align}
    {\hat P} = \left( \begin{array}{cccccc}
        0    &   0     &    0   &   0    &   0    &{\hat U}_{x} \\
        0    &   0     &    0   &    0   &{\hat N}_{x} &     0      \\
        0    &   0     &    0   &{\hat N}_{x} &   0    &    0       \\
        0    &   0     & ...   &   0    &   0    &    0       \\
        0    &{\hat N}_{x}  &   0    &  0     &  0     &   0        \\
{\hat U}_{x}  &     0   &    0   &   0    &   0    &       0    \\
\end{array} \right)
\end{align}
the invariance of the coupled system under rotation around the $x$-axis yields
\begin{align}
   {\hat P}^\dagger {\hat H}_S(k_x, k_y, N_z, \alpha, J, \lambda){\hat P} = {\hat H}_S(k_x, -k_y, N_z, \alpha, J, \lambda) \ .
\end{align}
This equation holds if
\begin{align}
   \begin{array}{c}
{\hat U}_x^{\dagger} {\hat V}_{b}  {\hat N}_{x} = {{\hat V}_{t}}^{\dagger} \\
{\hat N}_{x}^{\dagger} {\hat V}_{t}  {\hat U}_{x} = {{\hat V}_{b}}^{\dagger}
\end{array}
\end{align}
which is satisfied by choosing
\begin{eqnarray}
{\hat V}_{t} = V \left( \begin{array}{cccccccc}
1 & -1 & 0 & 0 & 0   & 0  & 0  & 0\\
0 & 0 & 1 & -1 &  0  & 0  & 0  & 0\\
0 & 0  & 0   & 0  &   -1 & 1 & 0 & 0\\
0 & 0  & 0   & 0  &   0 & 0 & -1 & 1
\end{array} \right); \quad
V_{b} = V \left( \begin{array}{cccc}
1 & 0 &0 &0\\
1 & 0 &0 &0\\
0 & 1 &0 &0\\
0 & 1 &0 &0\\
0 & 0& -1 & 0 \\
0 & 0& -1 & 0 \\
0 & 0& 0 & -1 \\
0 & 0& 0 & -1 \\
\end{array} \right) \ .
\end{eqnarray}
For the results shown in the main text, we consider only coupled systems with an odd number of layers, $N_z$. In this case, and to preserve the symmetry of the system, we take the hybridization matrices in layers $n=1, \cdots, (N_z-1)/2$ to be identical to ${\hat V}_t$, the hybridization matrices in layers $n=(N_z-1)/2+2,\cdots N_z$ to be identical to ${\hat V}_b$, and the hybridization matrix in layer $n=(N_z-1)/2+1$ to be zero.

\section{Evolution of spectral functions below $T_c$}

\begin{figure*}[htbp]
    \centering
    \includegraphics[width=6 in]{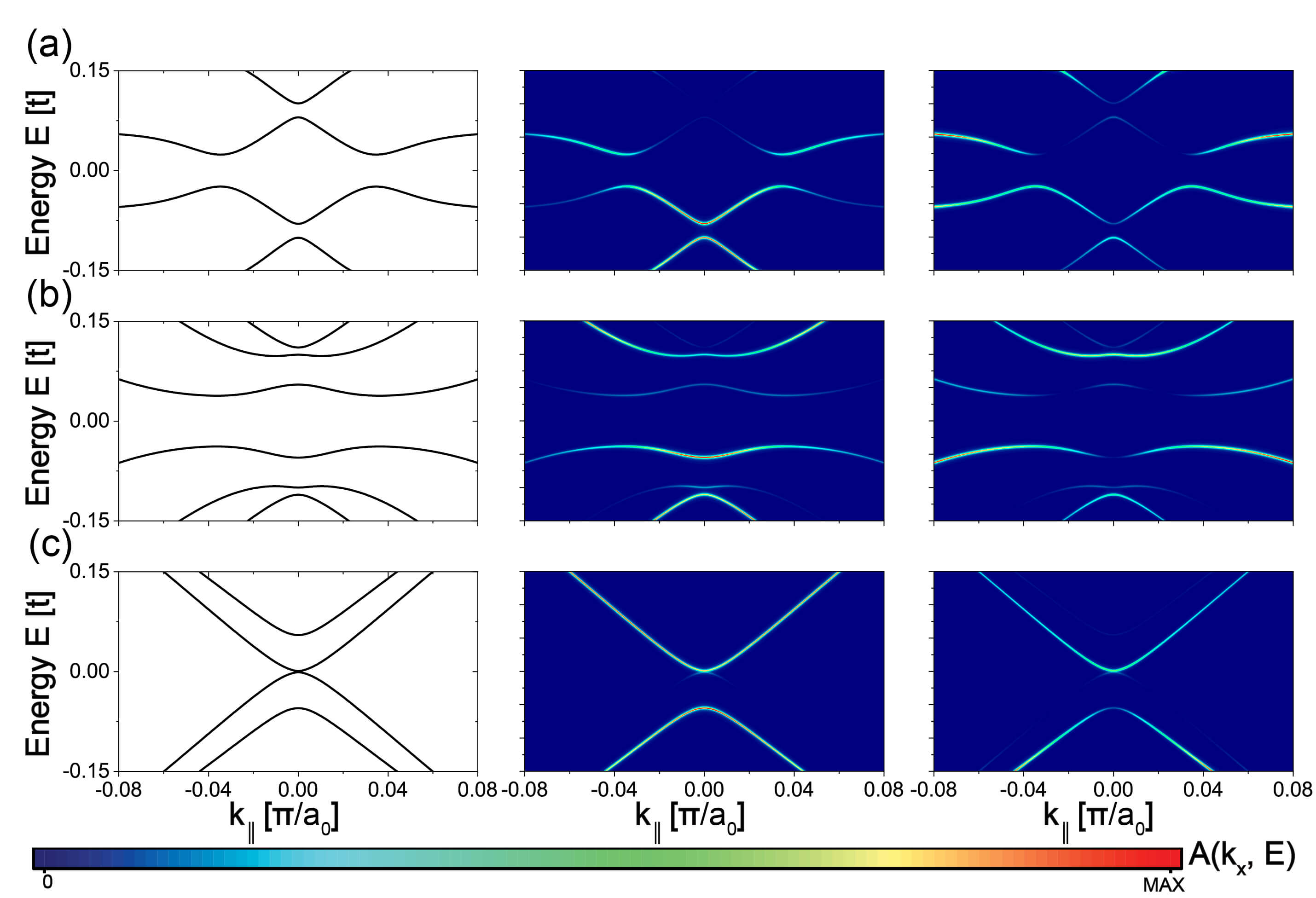}
    \caption{Evolution of the electronic structure (left column), (3DTI) $c$-electron (middle column) and (2DTSC) $f$-electron spectral functions (right column) below $T_c$ with increasing $JS$ as a function of $k_x$ for $k_y=0$: (a) $(JS,\mu_c)=(0.1,0.06)t$ (b) $(JS,\mu_c)=(0.2,0.04)t$, and (c) $(JS,\mu_c)=(0.45,0.04)t$. Parameters are $(V,\Delta_0,\alpha,\mu)=(0.07,0.1,0.1,3.8)t$.}
    \label{SIfig:fig2}
\end{figure*}

In Figs.2(a)-(c) of the main text, we presented the evolution of the electronic dispersion and of the (3DTI) $c$-electron spectral function with increasing $JS$ below $T_c$. In Fig.~\ref{SIfig:fig2}, we reproduce the results of Figs.~2(a)-(c) of the main text, together with the spectral function of the (2DTSC) $f$-electrons.
 It is interesting to note that neither the spectral function of the $c$-electrons, nor that of the $f$-electrons possesses any considerable weight in the negative energy branch of the band in which the gap closing occurs (see Figs.~\ref{SIfig:fig2}). Thus, independently of whether ARPES experiments probe the $c$- or $f$-electron orbitals \cite{Sobota2021}, the gap closing will be difficult to observe.

\section{Gap closings in the $(V,JS)$-plane}

As mentioned in the main text, a gap closing in the $(V,JS)$-plane occurs only at the $\Gamma$ point. Here, we demonstrate that there are no additional gap closings that occur at any other momenta in the Brillouin zone.
\begin{figure*}[htbp]
    \centering
    \includegraphics[width=6 in]{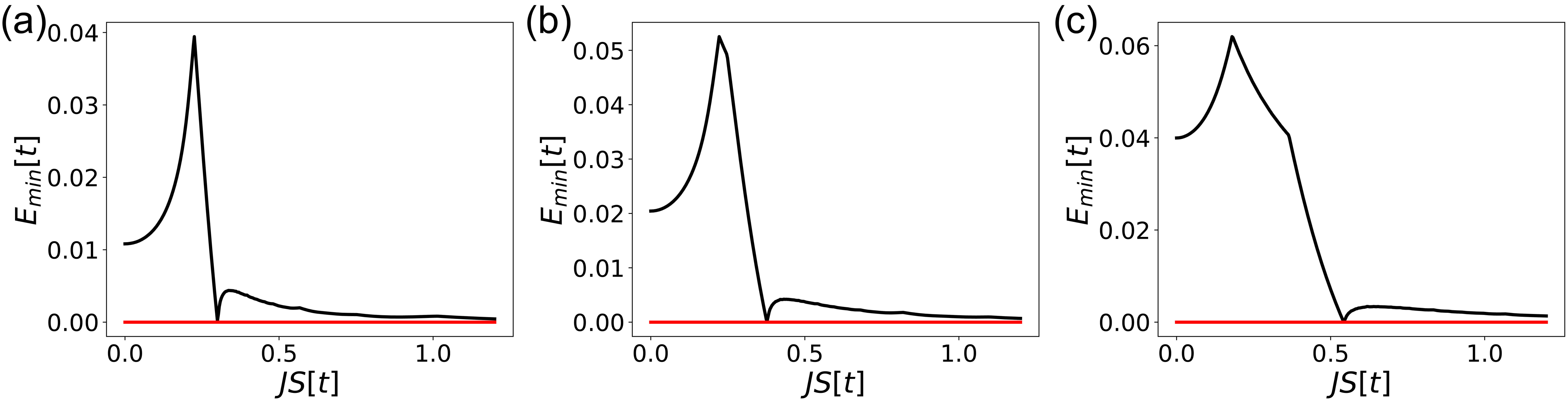}
    \caption{$E_{min}$ of the coupled 3DTI and 2DTSC system as a function of $JS$ for (a) $V=0.05t$, (b) $V=0.07t$, and (c) and $V=0.1t$.}
    \label{fig:gap}
\end{figure*}
To this end, we define the gap $E_{min}$ as the minimum positive energy for any momentum in the Brillouin zone, i.e., $E_{min} = \min_{{\bf k} \in BZ}(|E_{\bf k}|)$. Since determining $E_{min}$ for an extended range of parameters in the $(V,JS)$-plane is computationally very demanding, we consider only linecuts of $E_{min}$ as function of $JS$ for several values of the hybridization strength $V$, as shown in Fig.~\ref{fig:gap}. For all linecuts we considered, only a single gap closing occurs, which coincides with the gap closing at the $\Gamma$ point shown in the main text. Thus, we conclude that for the coupled 2DTSC and 3DTI systems, gap closings, which indicate a topological phase transition, occur only at the $\Gamma$-point.

\section{Relevance of bulk 2DTSC layers}

We argued in the main text, that the bulk 2DTSC layers that couple to the bulk 3DTI layers do not affect the surface electronic structure, or the topological surface phase. The reason for this is two-fold. First, the 2DTSC system itself exhibits a (quasi-)2D structure, with no direct coupling between the 2DTSC layers. This implies that the bulk 2DTSC layers cannot directly couple to the 2DTSC or 3DTI surface layers, only indirectly by hybridizing with the bulk 3DTI layers. This hybridization, however, is strongly suppressed since, secondly, the bulk 3DTI layers possesses a gap of $E_{g} = t$. Thus, the bulk 3DTI layers cannot hybridize with the low-energy states of the bulk 2DTSC layers.
Here, we demonstrate this explicitly by computing the electronic structure on the surface with and without the 2DTSC bulk bands, retaining in both cases the two 2DTSC layers that couple to the top and bottom surface layers of the 3DTI.
\begin{figure*}[htbp]
    \centering
    \includegraphics[width=6 in]{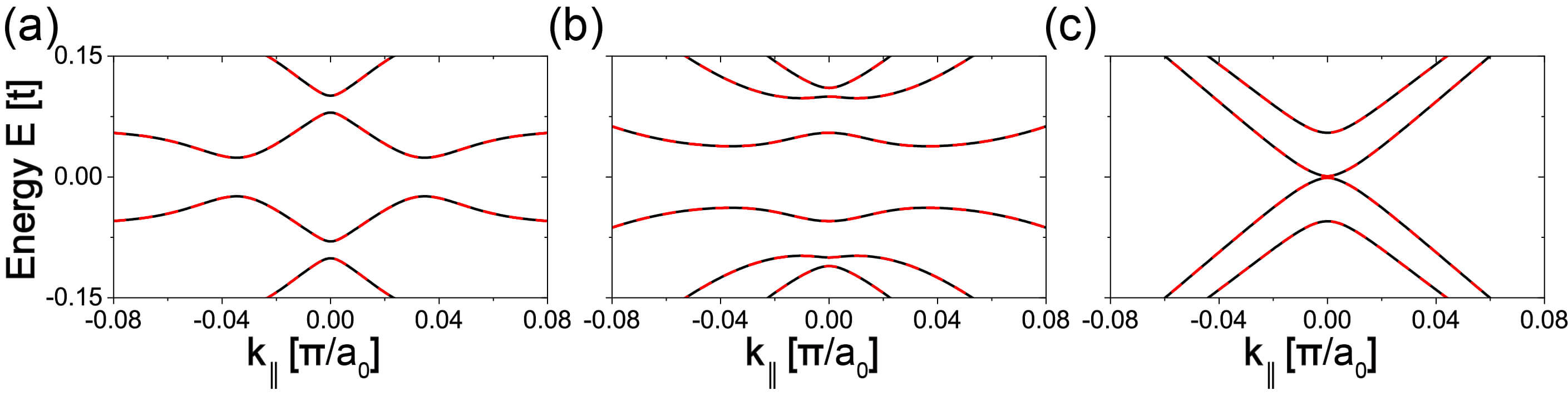}
    \caption{Evolution of the electronic structure below $T_c$ with increasing $JS$ as a function of $k_x$ for $k_y=0$: (a) $(JS,\mu_c)=(0.1,0.06)t$ (b) $(JS,\mu_c)=(0.2,0.04)t$, and (c) $(JS,\mu_c)=(0.45,0.04)t$. Black solid (red dashed) line represents the electronic structure when all surface and bulk 2DTSC layers (when only the surface 2DTSC layers) are included. Parameters are $(V,\Delta_0,\alpha,\mu)=(0.07,0.1,0.1,3.8)t$ and $N_z=5$.}
    \label{fig:fig_overlapped}
\end{figure*}
In Fig.~\ref{fig:fig_overlapped}, we present the resulting surface electronic structure for several values of $JS$ when surface and bulk 2DTSC layers are included (black solid line), and when only the two surface 2DTSC layers are included (red dashed line). These two results are essentially indistinguishable, implying that the bulk 2DTSC layers have no effects on the low-energy electronic structure of the surface.
Moreover, we find that the absence of the bulk 2DTSC layers does also not alter the gap map shown in Fig.2(d) of the main text, implying that the bulk 2DTSC layers do not affect the topological surface state of the hybridized system. To reduce the computationally complexity of our calculations, and to ensure that we can consider sufficiently large system sizes, we therefore neglect the bulk 2DTSC layers when studying the electronic structure in vortex cores and along domain walls.

\section{MZM in a vortex core}

For the calculation of the electronic structure around a vortex core, we consider a system that is finite in the $x$-, $y$-, and $z$-directions, with the system's length in these directions given by $l_i = N_i a_0$ $(i=x,y,z)$, with $a_0$ being the lattice constant. For the results shown in Fig.3 of the main text, we use two parameters sets characteristic of regions I and II. For region I [see Figs.3(a) and (c) of the main text], we present the $c$-electron LDOS (summed over both orbitals) for parameters  $(JS, \Delta_0, \mu_c, V, \alpha, \mu) = (0.0, 0.1, 0.06, 0.07, 0.1, 3.8)t$ and  system size $(N_x,N_y,N_z)=(81,81,7)$. For region II, we present [see Figs.3(b) and (d) of the main text] the $f$-electron LDOS with parameters $(JS, \Delta_0, \mu_c, V, \alpha, \mu) = (0.9, 0.6, 0.04, 0.07, 0.1, 3.8)t$ and system size $(N_x,N_y,N_z)=(151,151,3)$.

\begin{figure*}[htbp]
    \centering
    \includegraphics[width=8cm]{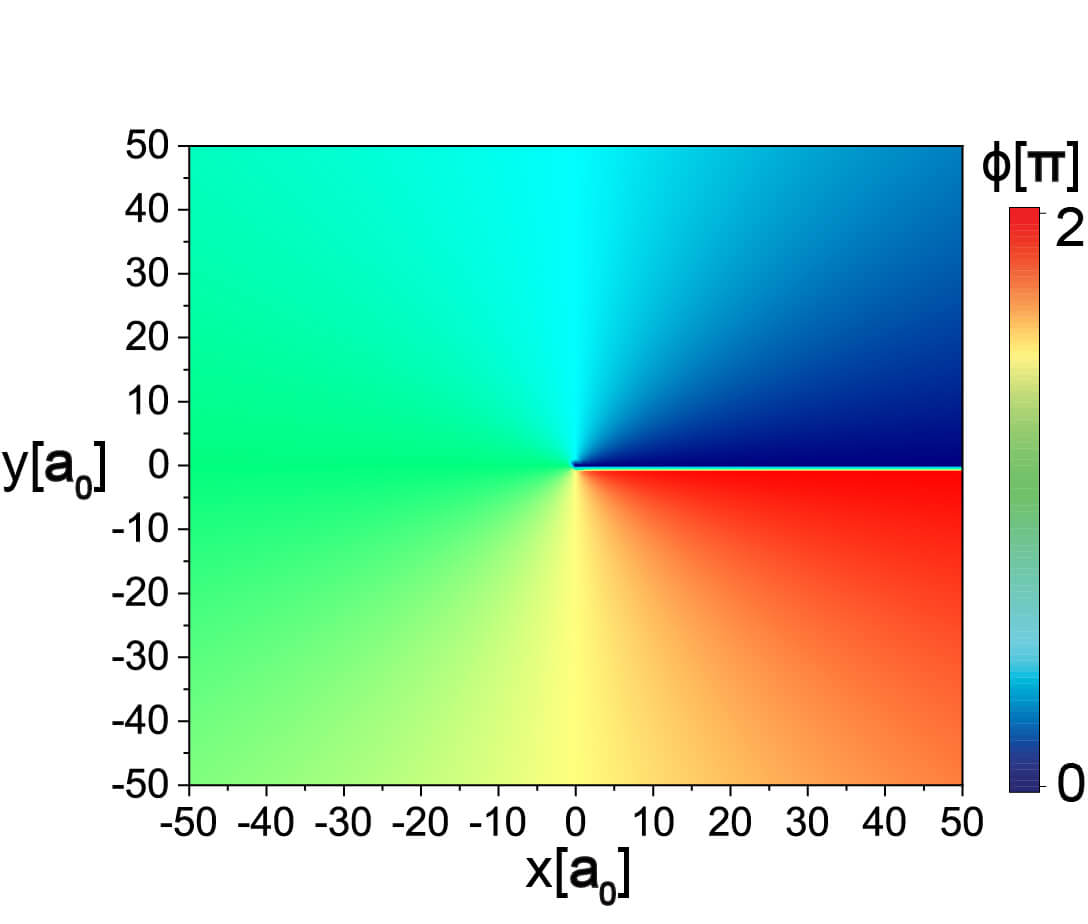}
    \caption{Phase angle $\phi({\bf r})$ as a function of position for the calculations shown in Fig.3 of the main text. }
    \label{fig:angle}
\end{figure*}
To simulate a magnetic vortex core in such a system, one typically introduces the magnetic field via a Peierls' substitution \cite{Mascot2022}, and computes the superconducting order parameter self-consistently. This approach, however, is computationally very demanding for the coupled 2DTSC and 3DTI systems, and the required system sizes. We therefore employ a simpler approach which has previously been used to study the emergence of MZMs in vortex cores \cite{Wu2021}. In this approach, one simulates a vortex core by imposing a phase-winding of the superconducting order parameter $\Delta ({\bf r}) =\Delta_0 e^{i \phi({\bf r})}$ around the vortex core. The spatial dependence of the phase $\phi({\bf r})$ is shown in Fig.~\ref{fig:angle}. As the magnetic flux penetrates the system from the bottom to the top surface, we assign the same superconducting phases on both surfaces. The LDOS in the vicinity of a vortex core for parameters characteristic for region I and II, are shown in Fig.3 of the main text.

We note that the small energy splitting of the MZMs shown in Fig.3 of the main text is the result of the finite size of the system, and the resulting hybridization between MZMs. Specifically, in region I, where the topological phase arises from the 3DTI mechanism, the energy splitting between the MZMs decreases with increasing number of layers $N_z$, i.e., with increasing separation between the two MZMs on the top and bottom surfaces. In contrast, in region II, the energy splitting decreases with increasing $N_x,N_y$, as the 2DTSC exhibits MZMs not only in a vortex core, but also at the edge of the system, as was previously discussed \cite{Rachel2017}.

\section{Majorana edge modes at domain walls}

In Fig.4 of the main text, we computed the electronic structure near PPDW and SDW in regions I and II using the parameters $(V, \alpha, \mu) = (0.07, 0.1, 3.8)t$. For region I, we used a characteristic parameter set given by $(JS, \Delta_0, \mu_c) = (0.1, 0.1, 0.06)t$, while for region II, we used $(JS, \Delta_0, \mu_c) = (0.9, 0.6, 0.04)t$. The system is implemented using periodic boundary conditions, such that it possesses 2 domain walls. The spectral functions are shown to the right of the domain walls, which are located between two columns of sites. To increase the gap in the system, we computed the electronic structure for the PPDW in region I (region II) using a scattering potential of $U_0=0.1$t. ($U_0=0.5$) [see Eq.(\ref{eq:scat})].

\begin{figure*}[htbp]
    \centering
    \includegraphics[width=17cm]{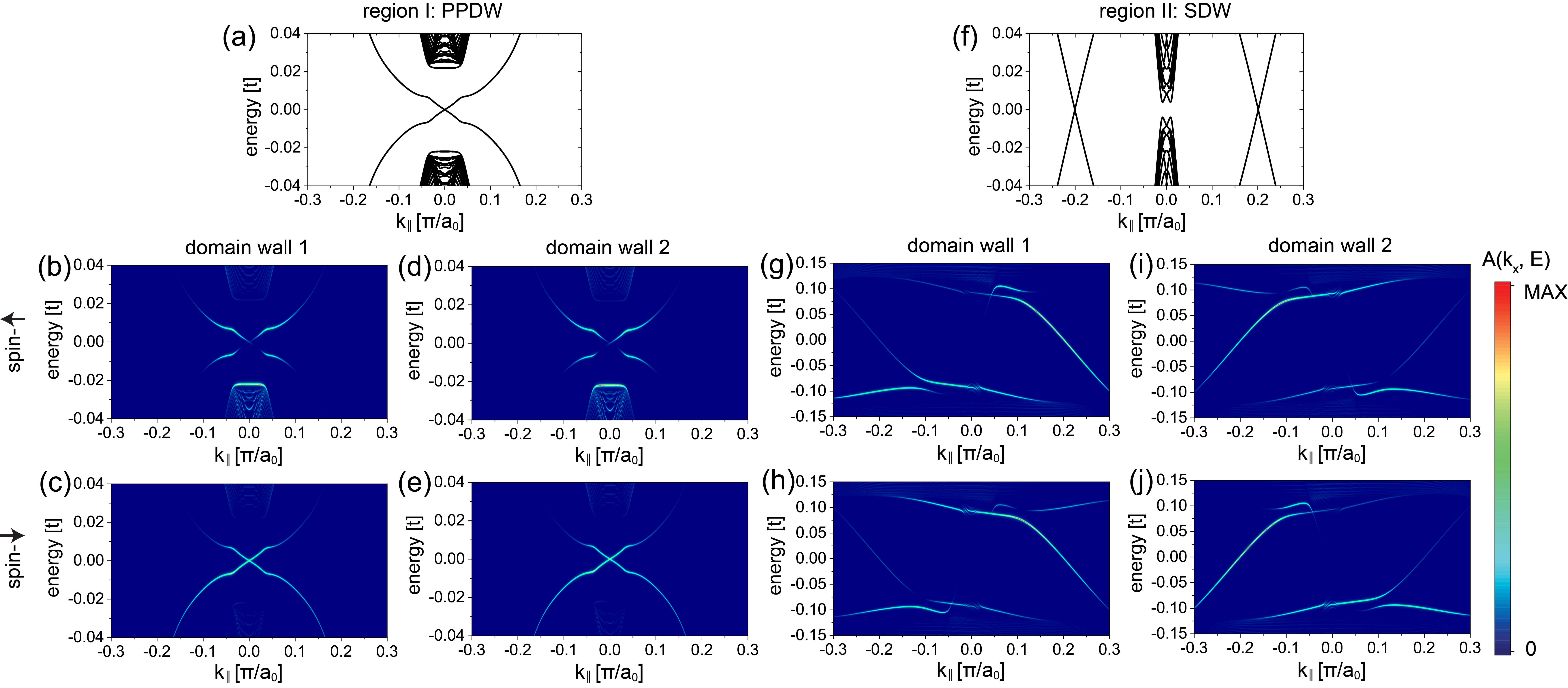}
    \caption{(a) Electronic dispersion as a function of momentum along a PPDW in region I. (b)-(e) Corresponding spin-resolved (3DTI) $c$-electron spectral functions to the right of domain wall 1 (left column) and domain wall 2 (right column). The spectral functions are summed over both $c$-electron orbitals, and parameters are $(V, \alpha, \mu, JS, \Delta_0, \mu_c, U_0) = (0.07, 0.1, 3.8, 0.1, 0.1, 0.06, 0.1)t$. (f)  Electronic dispersion as a function of momentum along a SDW in region II. (g)-(j) Corresponding spin-resolved (2DTSC) $f$-electron spectral functions to the right side of domain wall 1 (left column) and domain wall 2 (right column). Parameters are $(V, \alpha, \mu, JS, \Delta_0, \mu_c, U_0) = (0.07, 0.1, 3.8, 0.9, 0.6, 0.04, 0.0)t$, and $N_z=5$.}
    \label{fig:SI_all_walls}
\end{figure*}

Moreover, since the systems exhibits two domain walls, we can investigate the helical or chiral nature of the Majorana edge modes by considering the spin-resolved spectral functions at both domain walls. In Fig.~\ref{fig:SI_all_walls}(a) we present the the electronic structure shown in Fig.4 of the main text for the PPDW in region I, and in Figs.~\ref{fig:SI_all_walls}(b)-(e) the spin-resolved (3DTI) $c$-electron spectral functions at both domain walls. While the spectral functions exhibit a significant spin-dependence, they are identical at both domain walls, implying that the Majorana edge mode at a PPDW is neither helical nor chiral. In contrast, a comparison of the electronic structure for a SDW in region II, shown in Fig.~\ref{fig:SI_all_walls}(f), and the corresponding spin resolved (2DTSC) $f$-electron spectral functions shown in Figs.~\ref{fig:SI_all_walls}(g)-(j), clearly reveals the chiral nature of the Majorana edge mode.

Moreover, Majorana edge modes are topologically protected and thus robust against disorder effects. To demonstrate this robustness, we study the effects of disorder on the Majorana edge modes at a PPDW in region I [see Fig.4 (a) of the main text], and at a SDW in region II [see Fig.4 (c) of the main text]. To this end, we introduce a non-magnetic scattering potential along the domain wall on the two surfaces, defined via
\begin{align}
   {\hat H}_{scat}= U_0 \sum_{n=1,N_z} \sum_{{\bf R}} \sum_{{\bf k}_\parallel} \left( f_{n,{\bf k}_\parallel, {\bf R}, \sigma}^{\dagger} f_{n,{\bf k}_\parallel, {\bf R}, \sigma} +
   \sum_{a=1,2} c_{n,{\bf k}_\parallel, {\bf R}, a, \sigma}^{\dagger} c_{n,{\bf k}_\parallel,  {\bf R}, a, \sigma} \right)
   \label{eq:scat}
\end{align}
where $U_0$ is the scattering strength, $n$ is the layer index,  ${\bf R}$ denotes sites next to the domain wall (which is located between two lattice points), and ${\bf k}_\parallel$ is the momentum parallel to the domain wall.
\begin{figure*}[htbp]
    \centering
    \includegraphics[width=12cm]{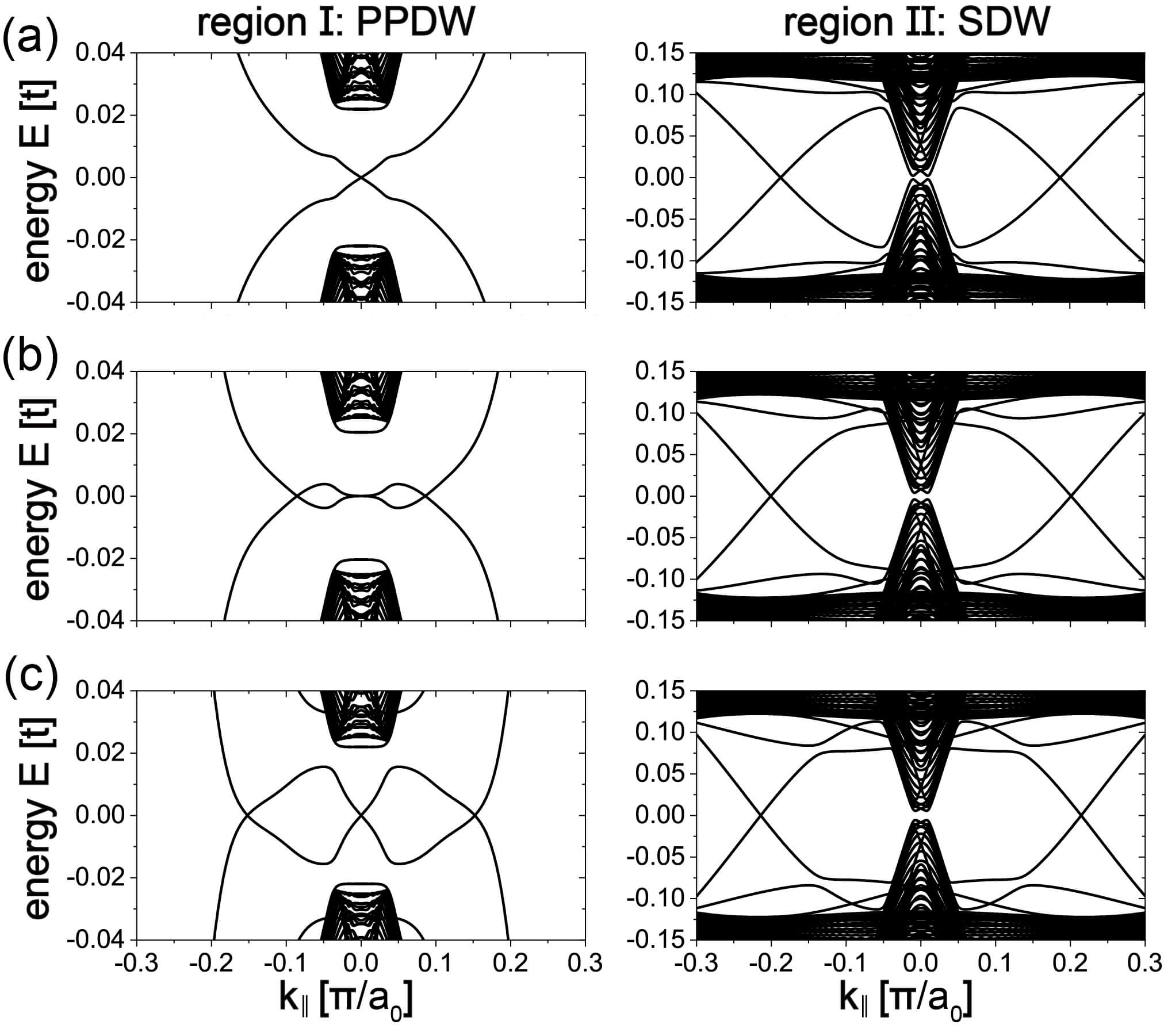}
    \caption{Electronic dispersion at domain walls in the presence of a non-magnetic scattering potential for a PPDW in region I (left column) and a SDW in region II (right column). (a) Electronic dispersion for a scattering potential of $U_0=0.1t$. (b) Electronic structures for zero scattering potential.  (c) Electronic structures for $U_0=-0.1t$. Parameters in region I are $(V, \alpha, \mu, JS, \Delta_0, \mu_c) = (0.07, 0.1, 3.8, 0.1, 0.1, 0.06)t$, while in region II we employed $(V, \alpha, \mu, JS, \Delta_0, \mu_c) = (0.07, 0.1, 3.8, 0.9, 0.6, 0.04)t$.}
    \label{fig:fig_disorder}
\end{figure*}
 In Fig.~\ref{fig:fig_disorder} we present the electronic dispersion for a PPDW in region I and a SDW in region II for different values of $U_0$. Note that the existence of Majorana edge modes, as well as the number of these modes, is unaffected by the scattering potential, as expected for topologically protected Majorana edge modes.

\section{Persistent supercurrents along domain walls}

The supercurrents flowing parallel to the domain wall possess two contributions, one each from a current flowing between the 2DTSC $f$-orbitals, and the 3DTI $c$-orbitals. There is no current flowing between the 2DTSC and 3DTI orbitals parallel to the domain wall as the hybridization is local, i.e., on-site.

The persistent supercurrent associated with the hopping of an electron from a site ${\bf r}$ to a nearest-neighbor site ${\bf r}+{{\bm \delta}}$ between 3DTI orbitals is given by
\begin{widetext}
\begin{align}
 I^{3DTI}_{{\bf r},  {\bf r}+{ {\bm \delta}}} &=
 -\frac{2e}{\hbar} \sum_{\sigma, \sigma'} \sum_{a,b= 1,2} \int \frac{d\omega}{2\pi}
  \text{Re} \left[
   \left(
    -t_{aa} \delta_{ab}\delta_{\sigma \sigma'}
    + i \lambda \left( {\bm \delta} \times {\bm \sigma}\right)^z_{\sigma \sigma'} \delta_{a{\bar b}}
   \right)
   g^{<}_{b,\sigma';a,\sigma}({\bf r}+{ {\bm \delta}}, {\bf r},\omega)
  \right] \ ,
 \label{eq:I1}
\end{align}
\end{widetext}
where  $-t_{aa}$ is the intra-orbital hopping amplitude between nearest neighbor sites, and $\lambda$ is the Rashba spin-orbit interaction in the 3DTI system, see Eq.(2) of the main text.
The persistent current between the 2DTSC orbitals is given by
\begin{widetext}
\begin{align}
 I^{2DTSC}_{{\bf r},  {\bf r}+{ {\bm \delta}}} &=
 -\frac{2e}{\hbar} \sum_{\sigma, \sigma'} \int \frac{d\omega}{2\pi}
  \text{Re} \left[
   \left(
    -t \delta_{\sigma \sigma'}
    + i \alpha \left( {\bm \delta} \times {\bm \sigma}\right)^z_{\sigma \sigma'}
   \right)
   g^{<}_{\sigma';\sigma}({\bf r}+{ {\bm \delta}}, {\bf r},\omega)
  \right] \ ,
 \label{eq:I2}
\end{align}
\end{widetext}
where $-t$ is the hopping amplitude between nearest neighbor sites, and $\alpha$ is the Rashba spin-orbit interaction in the 2DTSC system, see Eq.(1) of the main text. Moreover,
${g}^{<}_{\sigma,a;\sigma',b}({\bf r},{\bf r}+{ {\bm \delta}},\omega)$ [$g^{<}_{\sigma';\sigma}({\bf r}+{ {\bm \delta}}, {\bf r},\omega)$] are the $(\sigma,a;\sigma',b)$ [ $(\sigma;\sigma')$] elements in Nambu space of the lesser Green's function matrices in the 3DTI [2DTSC] system. To compute the lesser Greens functions, we rewrite the Hamiltonian of Eq.({\ref{eq:coupledH}}) for a ribbon geometry. Diagonalizing this Hamiltonian then allows us to compute the lesser Greens functions from the resulting eigenvectors and eigenvalues \cite{Mascot2022}.